\documentclass{PoS}

\usepackage{amsmath,amssymb,amscd,amsfonts,graphicx,bbm}

\rightline{\sffamily RBRC-778}\vspace*{-10mm}

\title{Improving chiral property of domain-wall fermions by reweighting method}

\ShortTitle{Improving chiral property of domain-wall fermions
            by reweighting method}

\author{\speaker{Tomomi Ishikawa}
        \thanks{Present address:
                Physics Department, University of Connecticut,
                Storrs, CT 06269-3046, USA}\\
        RIKEN BNL Research Center, Brookhaven National Laboratory,
        Upton, NY 11973, USA\\
        E-mail: \email{tomomi@quark.phy.bnl.gov}}

\author{Yasumichi Aoki\\
        RIKEN BNL Research Center, Brookhaven National Laboratory,
        Upton, NY 11973, USA\\
        E-mail: \email{yaoki@quark.phy.bnl.gov}}

\author{Taku Izubuchi\\
        RIKEN BNL Research Center, Brookhaven National Laboratory,
        Upton, NY 11973, USA\\
        Physics Department, Brookhaven National Laboratory,
        Upton, NY 11973, USA\\
        E-mail: \email{izubuchi@quark.phy.bnl.gov}}

\author{for the RBC and UKQCD Collaborations}

\abstract{
The reweighting method is applied to improve the chiral property
of domain-wall fermions.
One way to achieve this is to enlarge $L_s$, the size of fifth dimension,
which controls the size of the induced chiral symmetry breaking.
While this is a type of reweighting method for shifting
the action parameter, it seems non-trivial since this reweighting means
change of the five dimensional lattice volume.
In this report, we address issues in this direction of 
reweighting and evaluate its effectiveness.}

\FullConference{The XXVII International Symposium on Lattice Field Theory\\
		July 25-31, 2009\\
		Peking University, Beijing, China}

\begin{document}

\section{Introduction}

The domain-wall (DW) fermion is one realization of the chiral fermion
on the lattice~\cite{Kaplan:1992bt, Shamir:1993zy}.
The DW fermion is constructed by five-dimensional formulation and has
a parameter $L_s$, which is the size of fifth-dimension.
The chiral symmetry is realized in the $L_s\rightarrow\infty$ limit. 
In realistic simulations, however, we have to take $L_s$ finite;
then the symmetry is slightly broken.
The broken chiral symmetry causes the additive mass shift $m_{res}$,
which is called ``residual mass'' and captures the degree of the breaking.
In recent lattice QCD simulations,
we are going into the regime in which the quark mass is lighter and lighter
and the volume is larger and larger.
The computational resources required are becoming ever larger
in these situations, and then it is hard to take $L_s$ so large that
the chiral symmetry breaking is negligible.
Improving the chiral property of the DW fermion is a natural thought
in this circumstance.
One possible strategy for this is introducing a twisted mass term
to improve chiral symmetry, which suppresses a topology change
~\cite{Renfrew:2009wu, Fukaya:2006vs}.
Recently, reweighting techniques are becoming widely used
in QCD simulations.
Among them, the reweighting for the quark mass parameter is intensively
applied and seems effective~\cite{Hasenfratz:2008fg}.
Our aim of this work is to enlarge $L_s$ by reweighting to improve the
chiral property.
In this report, we discuss some techniques toward the reweighting and
its effectiveness.
We also perform some simple tests as an experiment using $N_f=2+1$
dynamical DW fermion configurations with volume
$L^3\times T\times L_s=16^3\times32\times8$:
\vspace*{-1mm}
\begin{eqnarray}
\mbox{Conf [A]} &:&
\beta=2.30 ~\mbox{(Iwasaki gluon)}, ~m_{ud}=0.04, ~m_s=0.04, ~M_5=1.8,
\label{EQ:CONF-A}\\
\mbox{Conf [B]} &:&
\beta=2.13 ~\mbox{(Iwasaki gluon)}, ~m_{ud}=0.02, ~m_s=0.04, ~M_5=1.8,
\label{EQ:CONF-B}
\end{eqnarray}
\vspace*{-3mm}
produced by RBC and UKQCD Collaborations~\cite{Antonio:2008zz}.

\section{Reweighting factor}

We consider to reweight configurations with $L_s=L_1$ to $L_2$
($L_1<L_2$).
In this work, we ignore the strange quark sector even though we use
$N_f=2+1$ dynamical configurations because we assume the
effect is small and the present work is at an experimental stage.
The reweighting factor can be simply written by
\begin{eqnarray}
w=\frac{\det\left[D_2^{\dagger}(m_f)D_2(m_f)\right]}
       {\det\left[D_1^{\dagger}(m_f)D_1(m_f)\right]}
  \frac{\det\left[D_1^{\dagger}(  1)D_1(  1)\right]}
       {\det\left[D_2^{\dagger}(  1)D_2(  1)\right]},
\label{EQ:reweighting_factor}
\end{eqnarray}
where $D_i(m_f)$ represents DW Dirac matrix with parameter set
$(L_s=L_i, m_f)$, where $m_f$ represents the bare DW fermion mass.
One of the interesting features of this reweighting is the existence of
the part with $m_f=1$ which is coming from Pauli-Villars (PV) field
of the DW formalism.
While the systems before and after the reweighting have different
five-dimensional volume, the effect coming from the PV sector
cancels out a large portion of the volume factor.

\section{Stochastic estimation of the reweighting factor}

In order to estimate the reweighting factor we use the stochastic
estimator with random Gaussian noise.
When we consider the determinant of a matrix $\Omega$,
its stochastic estimation can be expressed by
\begin{eqnarray}
\det\Omega
=\frac{\int{\cal D}\xi e^{-\xi^{\dagger}\Omega^{-1}\xi}}
{\int{\cal D}\xi e^{-\xi^{\dagger}\xi}}
=\langle e^{-\xi^{\dagger}(\Omega^{-1}-1)\xi}\rangle_{\xi}
=\langle e^{-H}\rangle_{\xi},
\label{EQ:stochastic}
\end{eqnarray}
where $\langle\cdots\rangle_{\xi}$ denotes the ensemble average over
complex random Gaussian noise vector $\xi$.
A naive estimation, however, might end in failure.
In this section, we explain several implementations used
in the reweighting.

\subsection{Canceling the fluctuations}

A naive way to estimate the reweighting factor $w$
~(\ref{EQ:reweighting_factor}) is to calculate each determinant separately
by using Eq.~(\ref{EQ:stochastic}).
This way, however, is not efficient, because each determinant can have
large fluctuations.
To reduce the large fluctuation, a determinant of the whole product of
Dirac operators is estimated using one Gaussian noise $\xi$. 
For the efficient sampling of $e^{-H}$, we choose hermitian operator
$\Omega=\phi^{\dagger}\phi$ in Eq.~(\ref{EQ:stochastic}) and thus
the reweighting factor becomes $w=\det\left[\phi^{\dagger}\phi\right]$.
In this work, we take an operator for $\phi$ as
\begin{eqnarray}
\phi={\cal D}_2(m_f)\frac{1}{{\cal D}_1(m_f)}
 {\cal D}_1(  1)\frac{1}{{\cal D}_2(  1)},
\label{EQ:phi_definition}
\end{eqnarray}
where a notation, ${\cal D}=\sqrt{D^{\dagger}D}$, is used,
and the square root is implemented by the rational approximation
~\cite{Clark:2006fx}.
While we can, of course, write down the $\phi$ without using the square root,
we use it on purpose for later convenience (See Sec.~\ref{SEC:Stochastic}).
At the mathematical level, different order of the matrices
in Eq.~(\ref{EQ:phi_definition}) provides exactly the same value of $w$.
In the stochastic evaluation with finite statistics, however,
they could give different value and we need to investigate the optimal choice.

\subsection{Breaking up determinants}
\label{SEC:Stochastic}

While the statistical average of $e^{-H}$ always converges the correct
determinant $\det\Omega$ in the infinite number of sampling,
the estimation deviates from the true value significantly
for finite statistics.
Moreover it is difficult to estimate the size of the error as we will see
in the next subsection.
This is due to the long tail of the asymmetric distribution of $e^{-H}$,
and small number of outliers dominate the average.
To avoid this obstacle, it is found to be efficient to break up the determinant
into many number of smaller pieces \cite{Hasenfratz:2008fg}
so that the effect from the outliers is suppressed.
One way for the breaking up is splitting the parameters that we want to
shift in the reweighting, for example, splitting the mass parameter
in the mass reweighting.
In our reweighting, this splitting can be done by dividing $L_2-L_1$:
\begin{eqnarray}
w=\det\Omega=
\det\Omega_{L_1\rightarrow l_2}\cdot\det\Omega_{l_2\rightarrow l_3}\cdots
\det\Omega_{l_{n-1}\rightarrow l_n}\cdot\det\Omega_{l_n\rightarrow L_2},
\end{eqnarray}
where $L_1<l_2<\cdots<l_n<L_2$.
Another possible way is to use the so-called $n^{th}$ root trick.
It is provided by a simple mathematical identity:
\begin{eqnarray}
w=\det\Omega=\left(\det\Omega^{1/n}\right)^n.
\end{eqnarray}
This splitting can be easily performed by using the rational approximation
for the Dirac matrices, and we implement it in our simulation code
as we explained before.
In this breakup, the magnitude of the fluctuations for each divided
estimator is roughly $1/n$ times smaller than original. 
Of course, a hybrid method combining these two ways of breaking-up the
determinants is possible.

\subsection{Numerical demonstration of the techniques in the reweighting}

Here we show some demonstrations for the $n^{th}$ root trick and
combining the determinants.
The reweighting shown here is that from $L_s=8$ to $10$,
and we used one configuration
(traj id $=1000$ in Conf [A]~(\ref{EQ:CONF-A})). 

\subsection*{\underline{$n^{th}$ root trick}}

We discuss the effectiveness of breaking up the determinant  using
the $n^{th}$ root trick.
We consider the reweighting factor $w$ ~(\ref{EQ:reweighting_factor}),
and we take the $n^{th}$ root of it:
\begin{eqnarray}
w=\left(\det\left[\phi_n^{\dagger}\phi_n\right]\right)^n
=\prod_{i=1}^n\langle
e^{-\xi_i^{\dagger}((\phi_n^{\dagger}\phi_n)^{-1}-1)\xi_i}
\rangle_{\xi_i}
=\prod_{i=1}^n\langle e^{-h}\rangle_{\xi_i},
\label{EQ:nth-root-w}
\end{eqnarray}
with
\begin{eqnarray}
\phi_n=         \sqrt[n]{{\cal D}_2(m_f)}
       \frac{1}{\sqrt[n]{{\cal D}_1(m_f)}}
                \sqrt[n]{{\cal D}_1(  1)}
       \frac{1}{\sqrt[n]{{\cal D}_2(  1)}}.
\end{eqnarray}
\begin{figure}[t]
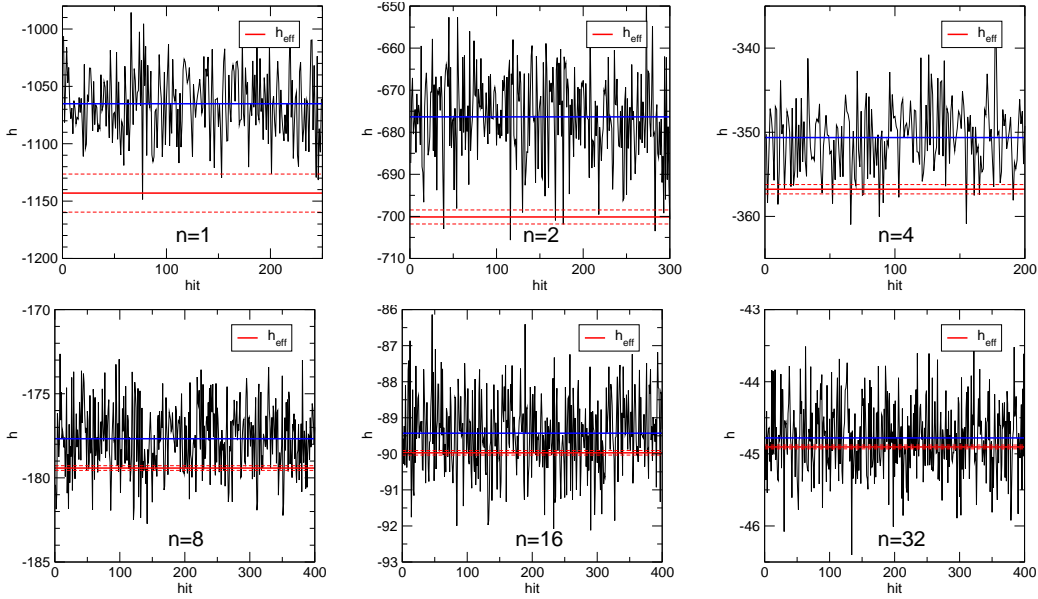

\begin{center}
\begin{tabular}{lcr}
\begin{minipage}{43mm}
 \includegraphics[scale=0.32, viewport = 0 0 380 340, clip]
{./Figures/heff_r01.eps}
\end{minipage}
&
\begin{minipage}{43mm}
\includegraphics[scale=0.32, viewport = 0 0 380 340, clip]
{./Figures/heff_r02.eps}
\end{minipage}
&
\begin{minipage}{43mm}
\includegraphics[scale=0.32, viewport = 0 0 380 340, clip]
{./Figures/heff_r04.eps}
\end{minipage}
\vspace*{2mm}\\
\begin{minipage}{43mm}
\includegraphics[scale=0.32, viewport = 0 0 380 340, clip]
{./Figures/heff_r08.eps}
\end{minipage}
&
\begin{minipage}{43mm}
\includegraphics[scale=0.32, viewport = 0 0 380 340, clip]
{./Figures/heff_r16.eps}
\end{minipage}
&
\begin{minipage}{43mm}
\includegraphics[scale=0.32, viewport = 0 0 380 340, clip]
{./Figures/heff_r32.eps}
\end{minipage}
\end{tabular}
\caption{History of $h$ in various $n^{th}$ root.}
\label{FIG:nth-root_trick:history}
\end{center}
\vspace*{-5mm}
\end{figure}
Fig.~\ref{FIG:nth-root_trick:history} shows the history of
$h$ in Eq.~(\ref{EQ:nth-root-w}).
$h_{\rm eff}$ in Fig.~\ref{FIG:nth-root_trick:history} is defined by
$e^{-h_{\rm eff}}=\langle e^{-h}\rangle_{\xi}$ and represented by red lines.
When $n=1$, that is, we do not impose the $n^{th}$ root trick,
the fluctuation of $h$ is so huge ($\sim O(10)$) that
the sampling is dominated by only one hit and tends to fail.
We find, however, that the magnitude of the fluctuation of $h$ behaves
as $\sim 1/n$ as $n$ increases, and the sampling gets close
to the situation in which $\langle h\rangle_{\xi}\sim h_{\rm eff}$
($\langle h\rangle_{\xi}$ is represented by blue lines).
Fig.~\ref{FIG:nth-root_trick:total_Heff} shows the total $H_{\rm eff}$
defined by $e^{-H_{\rm eff}}=\langle e^{-H}\rangle_{\xi}$.
We find that the case $n=1$ (no root) gives a quite misleading value, and
as $n$ increases, $H_{\rm eff}$ approaches an asymptotic value.
Note also the jackknife errors completely underestimate the
true error for small $n$.
The $16^{th}$ root seems sufficient to obtain the correct value
of the reweighting factor for this case.

\subsection*{\underline{Canceling the fluctuations}}

\begin{figure}[t]
\vspace*{-5mm}
\begin{center}
\begin{tabular}{lr}
\hspace*{-2mm}
\begin{minipage}{53mm}
\vspace*{+3mm}
\includegraphics[scale=0.38, viewport = 0 0 380 250, clip]
{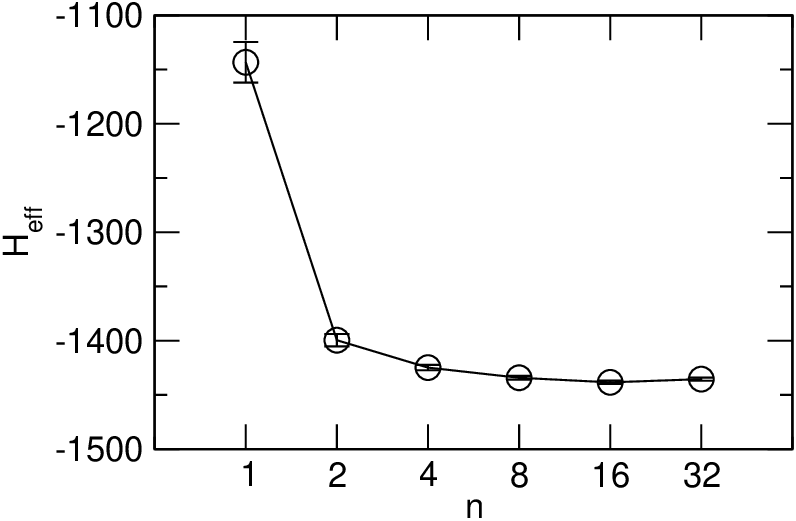}
\caption{$n$ dependence of total $H_{\rm eff}$.}
\label{FIG:nth-root_trick:total_Heff}
\end{minipage}
&
\begin{minipage}{97mm}
\begin{tabular}{lr}
\begin{minipage}{42mm}
\includegraphics[scale=0.32, viewport = 0 0 400 350, clip]
{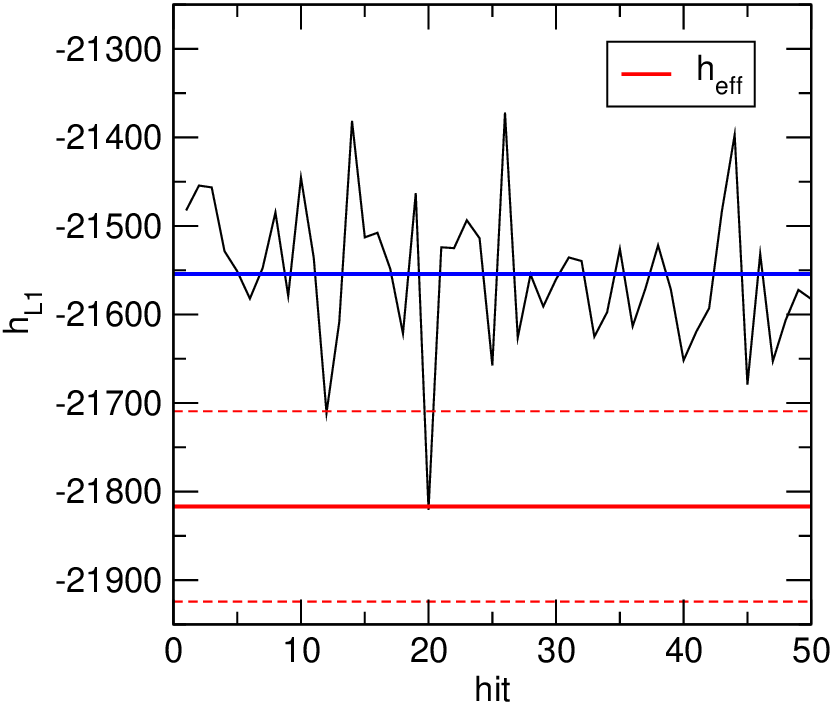}
\end{minipage}
&
\begin{minipage}{42mm}
\includegraphics[scale=0.32, viewport = 0 0 400 350, clip]
{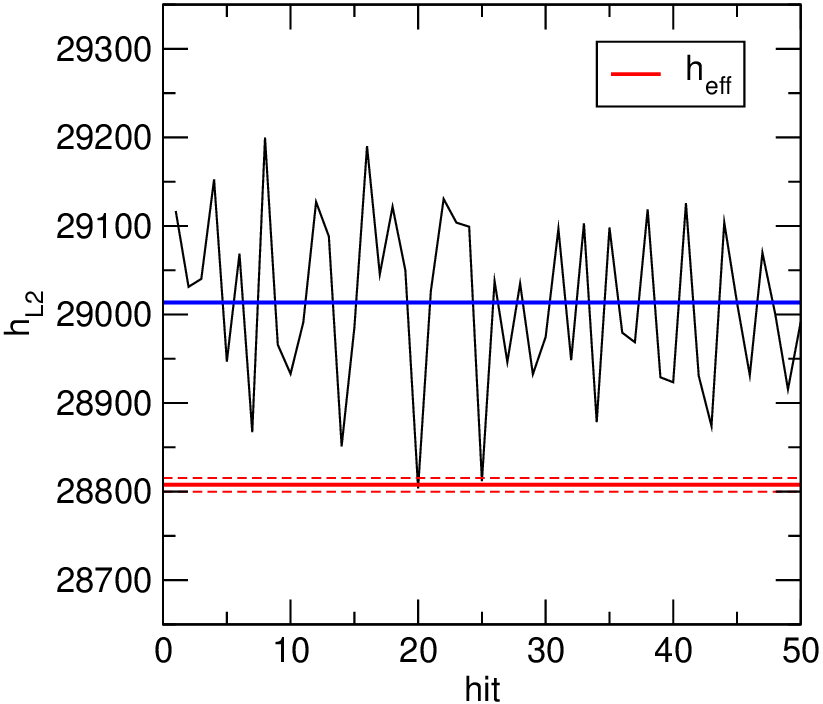}
\end{minipage}
\end{tabular}
\caption{History of $h_{\rm L_1}$ and $h_{\rm L_2}$.}
\label{FIG:cancelling_fluctuation}
\end{minipage}
\end{tabular}
\end{center}
\vspace*{-5mm}
\end{figure}

Here we discuss cancellation of the fluctuation.
As a test, we consider the $4^{th}$ root of the reweighting factor,
and we use two kinds of stochastic estimation:
\begin{eqnarray}
w^{1/4}&=&
\langle e^{-h_{\rm L_1}}\rangle_{\xi}\langle e^{-h_{\rm L_2}}\rangle_{\xi}
\;\;\;\;\;\;
{\rm or}
\;\;\;\;\;\;
\langle e^{-h_{\rm L_1L_2}}\rangle_{\xi},
\end{eqnarray}
where
\begin{eqnarray}
\langle e^{-h_{\rm L_2}}\rangle_{\xi}
=\det^{1/4}\left[\frac{{\cal D}_2(m_f)}{{\cal D}_2(1)}\right],\;
\langle e^{-h_{\rm L_1}}\rangle_{\xi}
=\det^{1/4}\left[\frac{{\cal D}_1(  1)}{{\cal D}_1(m_f)}\right],\;
\langle e^{-h_{\rm L_1L_2}}\rangle_{\xi}
=\det^{1/4}\left[\frac{{\cal D}_2(m_f)}{{\cal D}_1(m_f)}
                 \frac{{\cal D}_1(  1)}{{\cal D}_2(  1)}\right].
\end{eqnarray}
The history of $h_{\rm L_1L_2}$ is shown
in Fig.~\ref{FIG:nth-root_trick:history}
($n=4$) and of $h_{\rm L_1}$ and $h_{\rm L_2}$ are shown
in Fig.~\ref{FIG:cancelling_fluctuation}.
While magnitude of the fluctuation of the $h_{\rm L_1}$ and $h_{\rm L_2}$ is
around $200$,
that of the $h_{\rm L_1L_2}$ is largely reduced to around $10$.
It shows combining the determinants gives us a great advantage.

\section{Fluctuation of the reweighting factor 
         among different gauge configurations}

In this section we discuss the fluctuation of the reweighting factor $w$
among different gauge configurations.
Even when the correct $w$ for a given configuration
is obtained by the methods in previous section,
too large fluctuation of $w$ among different gauge configurations
could still ruin the precise estimation for the  final reweighed observable.
As an example, we show results of the reweighting from $L_s=8$ to $16$
using configurations of traj id $=1600 \sim 3500$ 
in Conf [B]~(\ref{EQ:CONF-B}). 

\subsection{Naive fluctuations}

\begin{figure}[t]
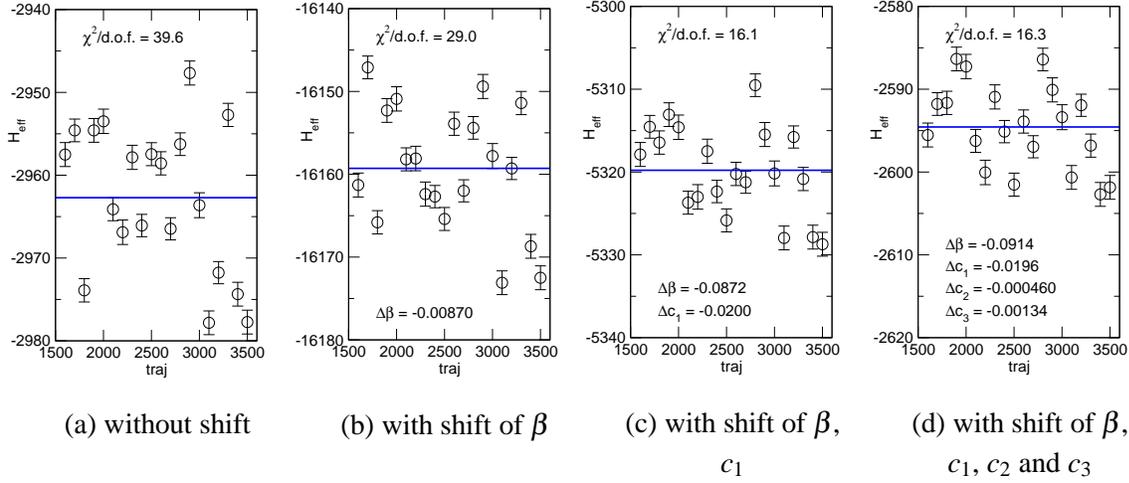

\vspace*{-5mm}
\begin{center}
\begin{tabular}{lccr}
\begin{minipage}{34mm}
\hspace*{-4mm}
\includegraphics[scale=0.35, viewport = 0 0 300 420, clip]
{./Figures/heff_vs_traj.eps}
\parbox[c]{34mm}{
\begin{center}(a) without shift \\ ~ \end{center}}
\end{minipage}
&
\begin{minipage}{34mm}
\hspace*{-5mm}
\includegraphics[scale=0.35, viewport = 0 0 300 420, clip]
{./Figures/heff_vs_traj_shifted_iwasaki.eps}
\parbox[c]{34mm}{
\begin{center}(b) with shift of $\beta$ \\ ~ \end{center}}
\end{minipage}
&
\begin{minipage}{34mm}
\hspace*{-5mm}
\includegraphics[scale=0.35, viewport = 0 0 300 420, clip]
{./Figures/heff_vs_traj_shifted_plaq_rect.eps}
\parbox[c]{34mm}{
\begin{center}(c) with shift of $\beta$, \\ $c_1$ \end{center}}
\end{minipage}
&
\begin{minipage}{34mm}
\hspace*{-5mm}
\includegraphics[scale=0.35, viewport = 0 0 300 420, clip]
{./Figures/heff_vs_traj_shifted_plaq_rect_chair_cube.eps}
\parbox[c]{34mm}{
\begin{center}(d) with shift of $\beta$, \\ $c_1$, $c_2$ and $c_3$ \end{center}}
\end{minipage}
\vspace*{-3mm}
\end{tabular}
\caption{$H_{\rm eff}$ in configuration to configuration.}
\label{FIG:Heff_conf_to_conf}
\end{center}
\vspace*{-4mm}
\end{figure}
Fig.~\ref{FIG:Heff_conf_to_conf}~(a) shows the obtained $H_{\rm eff}$
of each configurations.
The blue line in the figure represents a constant fit line
and its $\chi^2/\mbox{d.o.f.}$ is also put.
While the $H_{\rm eff}$'s themselves seem to be well determined,
their fluctuation among different configurations is quite large.
Then we conclude that the overlap between original and desired
configurations is so small that it is difficult to perform the reweighting
reliably for our parameter and volume.

\subsection{Shifting parameters}

Now we consider to compensate the reweighting factor for the large fluctuation
with shifting simulation parameters.
Here we consider a gluon action with standard combination:
\begin{eqnarray}
S_{\rm gluon}=-\beta\left(c_0[\mbox{plaquette}]+c_1[\mbox{rectangle}]
+c_2[\mbox{chair}]+c_3[\mbox{parallelogram}]\right),
\end{eqnarray}
with $c_0+8c_1+16c_2+8c_3=1$
($c_1=-0.331$ and $c_2=c_3=0$ on the original gluon configuration (Iwasaki glue)),
which means that we have a parameter space
$(L_s, \beta, c_1, c_2, c_3, m_f, M_5)$ in our simulation where $M_5$ denotes
DW height.
We, however, consider shifting parameters only in the gluon sector
$\beta$, $c_1$, $c_2$ and $c_3$ in the reweighting here.
\begin{figure}[t]
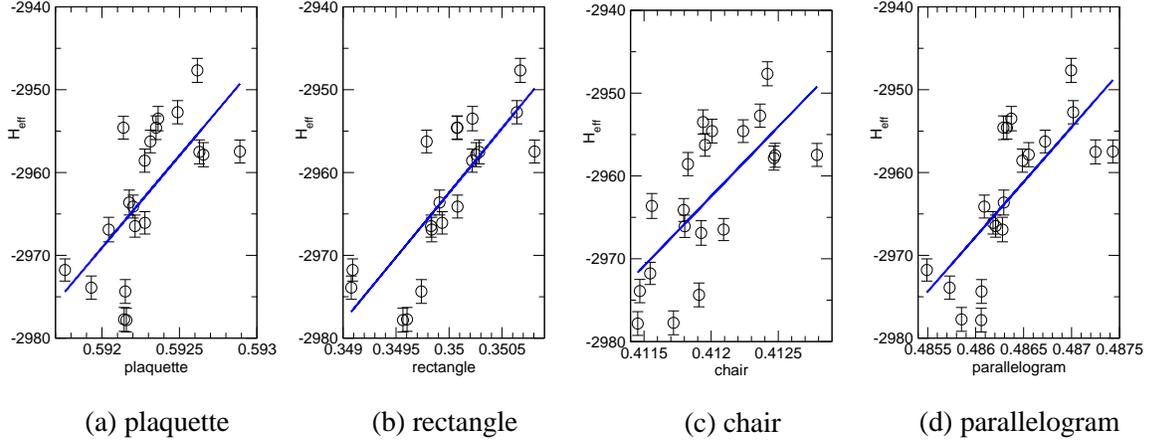

\begin{center}
\begin{tabular}{lccr}
\begin{minipage}{34mm}
\hspace*{-4mm}
\includegraphics[scale=0.35, viewport = 0 0 300 420, clip]
{./Figures/heff_vs_plaq.eps}
\parbox[c]{34mm}{
\begin{center}(a) plaquette \end{center}}
\end{minipage}
&
\begin{minipage}{34mm}
\hspace*{-5mm}
\includegraphics[scale=0.35, viewport = 0 0 300 420, clip]
{./Figures/heff_vs_rect.eps}
\parbox[c]{34mm}{
\begin{center}(b) rectangle \end{center}}
\end{minipage}
&
\begin{minipage}{34mm}
\hspace*{-5mm}
\includegraphics[scale=0.35, viewport = 0 0 300 420, clip]
{./Figures/heff_vs_chair.eps}
\parbox[c]{34mm}{
\begin{center}(c) chair \end{center}}
\end{minipage}
&
\begin{minipage}{34mm}
\hspace*{-5mm}
\includegraphics[scale=0.35, viewport = 0 0 300 420, clip]
{./Figures/heff_vs_cube.eps}
\parbox[c]{34mm}{
\begin{center}(d) parallelogram \end{center}}
\end{minipage}
\vspace*{-3mm}
\end{tabular}
\caption{Correlation between $H_{\rm eff}$ and various link loops.}
\label{FIG:Heff_correlation}
\end{center}
\vspace*{-5mm}
\end{figure}
Fig.~\ref{FIG:Heff_correlation} represents the correlation of $H_{\rm eff}$
with plaquette~(a), rectangle~(b), chair~(c) and parallelogram~(d).
Each quantity is likely to have a linear correlation with $H_{\rm eff}$
(the guide is shown by blue lines in each figures),
which means a shift of $\beta$ could help to reduce the fluctuation.
And although the trends of the figures seem to be the same,
they are still slightly different, which means we could use their combination
to reduce the fluctuation.
Figs.~\ref{FIG:Heff_conf_to_conf}~(b), (c) and (d) show the $H_{\rm eff}$
changed by optimised parameter shift which gives the smallest
$\chi^2/\mbox{d.o.f.}$ in the constant fit.
If we keep the value of the Iwasaki gluon parameter, and we only shift
$\beta$, the reduction of $\chi^2/\mbox{d.o.f.}$ is small.
(Fig.~\ref{FIG:Heff_conf_to_conf}~(b))
On the other hand, we can largely reduce $\chi^2/\mbox{d.o.f.}$
by shifting $c_1$.
(Fig.~\ref{FIG:Heff_conf_to_conf}~(c))
The shift of $c_2$ and $c_3$ does not contribute to the reduction so much.
(Fig.~\ref{FIG:Heff_conf_to_conf}~(d))
While the shifting the parameters in the gluon sector can contribute to the
reduction of the fluctuation, the fluctuations are so large that we are unable to perform the reweighting.
This indicates that we need to shift other parameters, that is,
$m_f$ and $M_5$.

\subsection{Balanced reweighting}

Besides shifting the parameters, an alternative idea is
{\it to balance more than one type of reweighting}.
While we investigate $N_f=2+1$ simulations,
the strange quark sector has not been accounted for in this study.
When the strange quark is included, we obtain an additional handle to control
the fluctuations.
If we consider chiral symmetry for the strange quark is less important
than that for the up and down quarks,
the fluctuation could be suppressed by enlarging the $L_s$ for the up and down
quarks while reducing $L_s$ for the strange, assuming large cancellation of
the fluctuation between the light and strange quark sector.
The up and down sector and the strange sector are, as it were,
well ``balanced''.
One of the other application of this concept might be the mass reweighting.
For example, in the $N_f=2+1$ simulation one could reduce the fluctuations
between the reweighting due to the shift of degenerate up and down sea quark
mass $m_{ud}$, and that of strange sea quark mass $m_s$ to the
first order in the mass shift $\Delta m$:
\begin{eqnarray}
w(m_{ud}\rightarrow m_{ud}-\Delta m;~ m_s\rightarrow m_s+2\alpha\Delta m)
=O(\Delta m^2),
\end{eqnarray}
where $\alpha$ is one at $m_{ud}=m_s$ and is decreasing function of
$(m_s-m_{ud})$, to be tuned by numerical calculations.

\section{Summary}

In this report, we discussed the possibility of the reweighting method
to enlarge $L_s$.
The reweighting factor itself can be calculated correctly by using
the stochastic method, dividing the determinant into many pieces.
In this study we used the $n^{th}$ root trick to control this,
and we found that it works well.
The problem lies on the fluctuation of the reweighting factor
among different gauge configurations, which is quite large.
Although we tried to make parameter shifts in the gluon sector to cure the
situation, it seems not enough.
We consider that additional parameter shifts, like $m_f$ and $M_5$, are needed
to suppress the fluctuation.
We are going to address this issue in the future.

\vspace*{-1mm}
\section*{Acknowledgments}
\vspace*{-1mm}
We acknowledge RIKEN, BNL and the U.S. DOE for providing the facilities
on which this work was performed.
We also thank Chulwoo Jung for helpful discussions.
This work is supported in part by JSPS KAKENHI 21540289.

\end{document}